# An architecture for the integration of different functional and structural plant models

Qinqin Long, Winfried Kurth
Institute of Computer Science
University of Göttingen
Göttingen, Germany
lqinqin@uni-goettingen.de

Christophe Pradal
CIRAD, UMR AGAP
INRIA, Zenith
University of Montpellier
Montpellier, France

Vincent Migault, Benoît Pallas
INRA, UMR AGAP
Montpellier, France

## ABSTRACT
Plant scientists use Functional Structural Plant Models (FSPMs) to model plant systems within a limited space-time range. To allow FSPMs to abstract complex plant systems beyond a single model's limitation, an integration that compounds different FSPMs could be a possible solution. However, the integration involves many technical dimensions and a generic software infrastructure for all integration cases is not possible. In this paper, we analyze the requirements of the integration with all the technical dimensions. Instead of an infrastructure, we propose a generic architecture with specific process-related components as a logical level solution by combining an ETL (Extract, Transform and Load) based sub architecture and a C/S (Client/Server) based sub architecture. This allows the integration of different FSP models hosted on the same and different FSP modeling platforms in a flexible way. We demonstrate the usability of the architecture by the implementation of a full infrastructure for the integration of two specific FSPMs, and we illustrate the effectiveness of the infrastructure by several integrative tests.

## 1. INTRODUCTION
With the development of computer science, especially computer graphics and relevant hardware and software technologies in recent decades, computer models that describe the function and structure of plants have emerged and been developed rapidly in plant research [1]. With or without taking environmental factors into account, these "Functional Structural Plant Models" (FSPMs) aggregate one or more functional simulators describing different physiological functions and a structural simulator describing the laws of plant structure change to digitally reproduce the interactions between structures and functions of plants along a part (or the whole) of their life cycle. The current mainstream approach of FSP modeling is platform based. The FSP modeling platform plays a role for FSP modeling similar to the role of the development kit for application development, e.g., JDK for Java applications. By providing crucial tools for describing plant systems, the platform is more of a domain-specific infrastructure than just a general development kit. Usually these includes a specific graphics library, a particular modeling formalism built upon a special modeling language with tailored operators and a FSP data model mostly detailed from a general data model (i.e. property graph), some useful components such as 3D viewers and "default" simulators that abstract general functional and structural processes of plants. By this approach, FSPMs are developed and executed on a given platform. As the modeling platform hides all computer-related technical details, plant scientists can thus use the tools provided by the platform transparently to build an FSPM in much shorter time and focus on their own specialty rather than on unfamiliar technologies.

Like all models, FSPMs abstract and simplify only a finite range of plants to a finite extent due to various constraints such as the available resources. It is technically impossible to model all physiological and environmental aspects of large complex botanical systems with many species by a single FSPM. To model complex botanical systems for a wide range of plants to a considerable extent, the capability of integration of different FSPMs is desired. "Integration" here means to build a compound model with synergy between existing models. Two requirements need to be met to achieve this goal. (1) The ability of information exchange between different FSPMs, and features to automatically interpret the exchanged information meaningfully and accurately to produce useful results as defined by the integrator, i.e., syntactic and semantic interoperability of information. (2) Simulators of different FSPMs must be synergistically operated to meet the objective of the integration, i.e., compliance of synergy of simulators.

In detail, the integration involves different technical dimensions. The first is the platform/model dimension. At platform level, plant information produced by FSPMs based on the same modeling platform shares the same syntax and semantics. Hence these FSPMs can use the same platform-level integrating infrastructure, i.e., processes for the platform-level interoperability of information. At model level, both information and simulator of a particular model have its own specific syntax and semantics, hence every FSPM has its unique model-level integrating infrastructure, i.e., processes for model-level interoperability of information and for synergy of simulators. The second is the syntax/semantics dimension. Information involved in FSPMs includes plant and environmental information. Both consist of data organized in syntactic structure with given semantics. Data need to be exchanged with relevant semantics to enable the simulation by the receiving FSPMs. The third is the dependent/independent dimension. Because plant elements are biologically dependent on and interact with each other, FSPMs compute the FSP data of one plant element by taking into account inputs from one or more other plant element. The plant information produced over one simulation step thus needs to be exchanged as a complete piece with consistent semantics in different syntax. In contrast, the different environmental information is normally considered independent from each other; hence, it does not need to be exchanged as a complete piece. The fourth is the topology/geometry dimension. Being a part of plant information, structural information includes topology and geometry. In detail, topology denotes the adjacency relationships between a plant element and its neighbors, whilst geometry denotes the location and orientation of a shape presenting a plant element and the geometric transformation between the plant element and its parents (local transformation) or the root (global transformation). The fifth is the internal/external dimension. The FSP data in plant information captures properties of the plant itself, i.e., internal data. In contrast, the data in environmental information is external. As FSPMs focus on small spatial scale modeling, the evolution of internal data is most often assumed to have no feedback on the external data, and the same external data is applicable for all involved virtual plants. The sixth is the non-retroactive/retroactive dimension. A non-retroactive integration denotes a situation in which the targeting FSPMs do not send the updated plant information back to the source FSPM. In contrast, a retroactive integration describes the case where the target FSPMs send the updated plant information back to the source FSPM and lets the source FSPM takes into account data on updated properties when computing new plant information.

Before FSPMs can be integrated, some preparations need to be carried out. One is plant property preparation. Similar to databases where different data fields characterize different properties of an object, different FSPMs originally organize data characterizing plant property information in different data field sets. However, the simulation of integrated FSPMs requires plant information with data fields from both source and target FSPMs. As the original plant information from the source FSPM does not contain data fields needed by the target FSPM, hence, the data fields defined in the target FSPMs need to be added to the data field set of the source FSPMs. The other is simulator preparation. Originally, simulators of an FSPM update plant information by computing new data of a data field using old data of relevant data fields defined in the FSPM itself. However, to compute new data of data fields defined in the source FSPMs in case of retroactive integration, the computation also takes data from data fields defined in the target FSPMs as inputs. Hence, simulators of the source FSPM need to be adjusted.

To integrate prepared FSPMs, syntactic and semantic interoperability of plant information between prepared FSPMs have to be enabled at both platform and model levels. At platform level, different FSP data models and graphics libraries are provided to capture plant function and structure. The data models define the syntaxes organizing individual data elements and the semantic relationships between data elements. The graphics libraries define the syntaxes (i.e. type signatures) and semantics (i.e. geometric meaning) of graphic types. Various plant data models and graphics libraries with different syntaxes and semantics are applied in FSP modeling practice, e.g. the RGG-based graph [2] and its IMP3D library provided by the platform GroIMP [3] [4] and the Multiscale Tree Graph (MTG) [5] and its PlantGL library [6] provided by the platform OpenAlea [7][8]. At model level, different FSPMs define their own syntaxes to organize data items (i.e. data fields) of each individual data element with specific given semantics. Different environmental data field sets defined in diverse syntaxes with different semantics can occur, e.g., to represent temperature, one FSPM can use "double" type representing Celsius degrees, another FSPM uses "float" type representing Fahrenheit degrees. Moreover, different multiscale structures are defined to suit different modeling cases syntactically. For example, FSPMs based on OpenAlea often apply a multiscale structure with metamers combining several elementary geometric objects as topological nodes at its finest scale. FSPMs based on GroIMP often apply a multiscale structure with one elementary geometric object as one topological node at its finest scale.

To integrate prepared FSPMs, the synergy of simulators of prepared FSPMs at model level need to be enabled as well. Simulators are algorithms capturing particular botanical knowledge with inputs and outputs. To enable the synergy, the outputs from simulators of the source FSPM need to be transformed to be the inputs to simulators of the target FSPMs. At the same time, the interactions between simulators of different FSPMs need to meet the objective of the integration fully.

Practically, the integration requires a distributed system in which different FSPMs are presented on FSP modeling platforms with a middleware that can synergize them with one another over a communication network. The middleware is in the middle of the distributed system for supporting the synergistic execution of distributed FSPMs. Evidently, our objective is to provide architecture for the distributed system to guide and facilitate the development of the middleware. To archive this, we have investigated solutions for the two integration requirements and designed a sub architecture which contains solution processes for each requirement. By combining the two sub architectures, we propose a full architecture for the integration of different FSPMs. We also demonstrate the usage of the architecture by the implementation of a full infrastructure for the integration of two specific FSPMs based on the platform GroIMP and OpenAlea respectively.

## 2. ARCHITECTURE FOR THE INTEGRATION OF DIFFERENT FSPMS
## 2.1 Sub architecture for syntactic and semantic interoperability of information

The interoperability of plant and environmental information includes the interoperability at both platform and model level. At platform level, it is about to make plant information using different data models and graphics libraries exchangeable and interpretable. To enable this, the syntaxes of information need to be transformed to carry the same semantics. The task is similar to the requirement of data integration, which deals with combining data in different syntaxes. Hence the approaches for data integration can be adapted to be the sub architecture for the interoperability of information. One straightforward approach is ETL (Extract, Transform, and Load) in data warehousing [9] [10].

In this approach, data is loaded into a data warehouse by going through three processes. (1) The extraction of data from one or more sources. (2) The transformations of the data, e.g. translating "1" in the source system to "M" in the target system for representing male. (3) The loading of the transformed data into specific target systems or file formats. This approach offers an architecture (i.e. ETL) with disassembled process units, thus it supports parallel processing. However, as an integral entity combining three functions, ETL is tightly coupled with specific source and target data models, i.e., a programming tool that implements ETL architecture is only applicable for particular source and target data models. Hence, it provides a dedicated information-exchange channel. An adapted approach with a unified mediating data model can make it more flexible. In this way, the information is exchanged between source and target FSPMs via the mediating data model. Two different ETLs are thus needed for each exchange direction. As the mediating data model is unified, processes that load to (or extract from) it are generic for any source and target models. Moreover, the mediating data model makes it possible for every potential target FSPM to load and process the same information, i.e., to run parallel simulations of different physiological processes within the same plant. We have designed a logical data exchange model, the Exchange Graph (EG) [11] as the mediating data model. As illustrated in Figure 1, in this model, basic properties are defined for nodes and edges, and a single root node that is directly or indirectly connected to all the other nodes provides the only entry point of the graph. The logical data exchange model can be detailed to physical data exchange models to allow the information exchange through implemented integrating infrastructure.

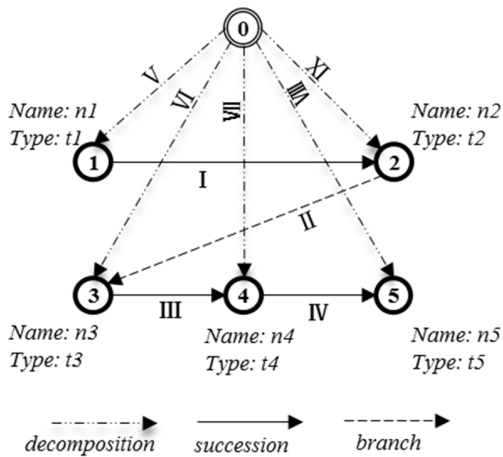

**Figure 1. An example of Exchange Graph.**

To enable the interoperability of information at platform level, ETL processes have to be defined according to the data models and graphics libraries of source and target FSPMs.

For the extracting and loading processes, the intra-scale structure is defined at the platform level as the basic part of the information model, hence the extraction and loading at platform level mostly concern the intra-scale structure, unlike in data warehousing where only data of primitive type are extracted and loaded. The interoperability of information usually requires more extracting and loading data of composite type, e.g., graphics types.

For the transforming process, several sub processes are necessary to meet the requirements of the integration. (1) Syntactic and semantic transformation of topology of data elements, e.g., generation of an edge in the target graph between corresponding source and destination nodes, and assignment of an edge type according to the edge type in the source graph, e.g., assignment of the "refinement" type available in the target FSP data model to generate an edge according to the "decomposition" type in the source graph. Essentially, the topology here concerns the structure with equivalent systems of scales (spatial resolutions). (2) Semantic transformation of the geometry of data elements. This may include a sub process that transforms geometric transformations between local and global. This sub-process is then essentially converting geometric relationships between nodes and is based on graph traversal. To avoid double running of graph traversal, this process is better run with the corresponding extracting process. Another sub process performs syntactic and semantic transformations of shape instances, e.g., transforms a signature with argument values of "Parallelogram" type to a signature with argument values of "TriangleSet" type. To allow this sub process, a "dictionary" to "translate" types from the graphics library used in the source FSPM to types in the graphics library used in the target FSPMs will be necessary.

To enable the interoperability of information at model level, ETL processes have to be defined also for functional and environmental information.

The functional and environmental information is usually specified in a FSPM using primitive types (e.g. float or integer). Thus extracting and loading processes are not really needed. The transformation process includes two sub processes. (1) Syntactic and semantic translation of coded values or derivation of new calculated values for functional or environmental data fields (e.g., float_ Fahrenheit = (float) (double_ Celsius * 1.8 + 32)). (2) Syntactic and semantic transformation of different systems of scales, for multiscale structures (e.g., decomposition of a scale with metamers as nodes to a new scale with elementary geometric objects as nodes) defined in different FSPMs. In most cases, simulators are applied to the finest scale of the multiscale graph. Thus the received structure needs to be transformed into a new structure with the finest scale on which the target simulator can be applied.

Figure 2 illustrates the sub architecture that combines the platform and model level ETL processes. It is clear that the infrastructure developed for integration will have a part that is general for FSPMs based on the same platform and a part that is only applicable for the particular source and target FSPMs.

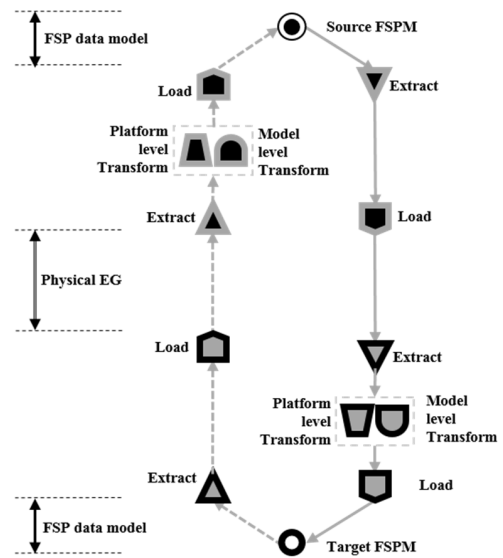

**Figure 2. Sub architecture for syntactic and semantic interoperability of information.**

## 2.2 Sub architecture for synergy of simulators

The integration of different FSPMs combines biological processes captured by different simulators at model level. It shall serve to allow one process to operate the other and has to be fully compliant with the specific integration objective. Due to the lack of a solution for automatic simulator recognition and interoperation management, the interoperation needs to be manually managed. Moreover, as a part of the middleware of a distributed system, processes for sending/receiving plant and environmental information and operating appropriate simulators with the right input list are also needed. Based on the nature of the requirements, we designed a sub architecture adapting the Client-Server (C/S) architecture [12] [13]. In the classic C/S architecture, one server normally receives requests from several clients. In our architecture, there can be more than one server involved and each one receives a request from the same client. The target FSPMs may be developed and managed by others while the source FSPM is usually developed and managed by the integrator himself. The server process of a target FSPM needs to be manually launched. The client process of the source FSPM also needs to be manually launched to execute the source FSPM and to start to send information to the target FSPM. The integrator assures that the combination of the simulators meets the expectations. Through these manual steps, the user of the integrated FSPMs manually manages the interoperation. A specific characteristic of the sub architecture is that there is only one client but one or more servers. The reason is that the integrated FSPMs always operate on one copy of data that is originally generated by the source FSPM with the initial plant structure. It is not logical to have more than one plant structure from different FSPMs during the simulation of the integrated FSPMs. Moreover, an integrator, who is normally also the developer of one FSPM to be integrated, only has full control of the FSPM developed by him, thus the role of source is fixed to a specific FSPM and there is a single client process. Additionally, the communication messages are different between the retroactive and non-retroactive integration. If the integration is non-retroactive, the response message contains only status information of the server FSPM execution. Otherwise, it contains also the updated plant information.

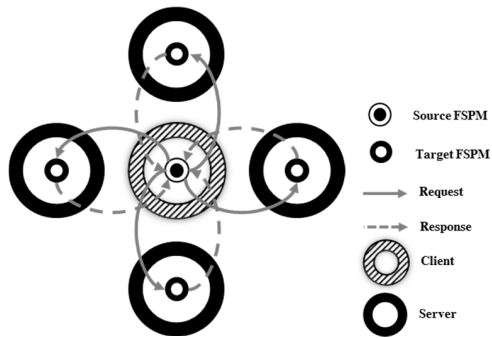

**Figure 3. Sub architecture for synergy of simulators.**

Figure 3 illustrates the adapted client-server sub architecture for synergy of simulators of different FSPMs. The FSPM in the center is a source model and the others are target models. Between each pair of source and target models, there is a client commonly used for all information exchange of the source FSPM and a dedicated server for information exchange only for a particular target FSPM.

## 2.3 Full architecture for the integration of different FSPMs

We propose that the data is better transformed during the movement from the mediating model to the target model. Since we want data in the mediating model to be reusable, the data thus needs to be transformed into a target data model only when the data in the mediating model is being moved to the target model. Therefore, there is no transformation process between the extracting and loading processes between source and mediating data model, and these two processes can be merged into one process that exports data from the source data model to the mediating data model.

As shown in figure 4, we get the complete architecture for the integration of different FSPMs by combining the two sub architectures.

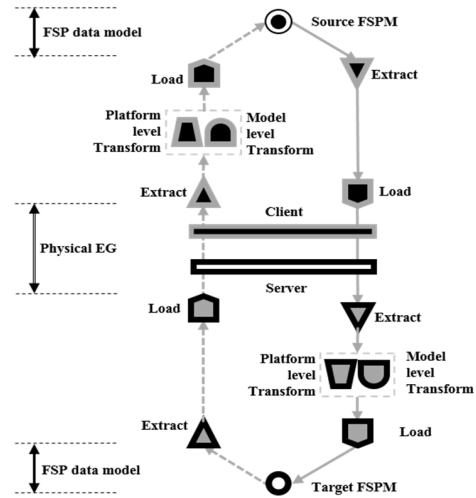

**(a) Pairwise view of the architecture.**

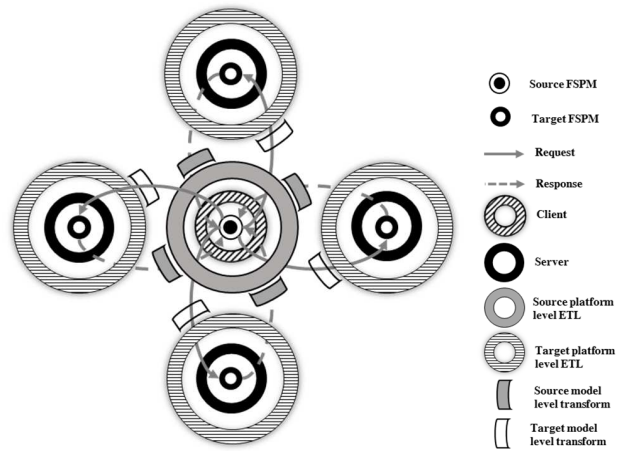

**(b) One-to-many view of the architecture.**

**Figure 4. Full Architecture for the integration of different FSPMs.**

## 3. MODEL INTEGRATION BY APPLYING THE ARCHITECTURE

### 3.1 Design and implementation for the integration of GroIMP and OpenAlea based models

In a current joint project [14], we want to provide an infrastructure to enable the integration of a GroIMP based FSPM [15] that simulates water pressure within the xylem of an apple tree and an OpenAlea based FSPM (i.e. MAppleT [16]) that simulates apple tree growth. For this purpose, we detailed the EG to an XML based physical Exchange Graph, XEG, as the physical mediating model. Figure 5 illustrates a plant with three elements and the code of its corresponding XEG. Functional and structural information of each plant element is assigned to relevant data fields, i.e., properties of corresponding node in the graph. Nodes are connected with each other through typed edges.

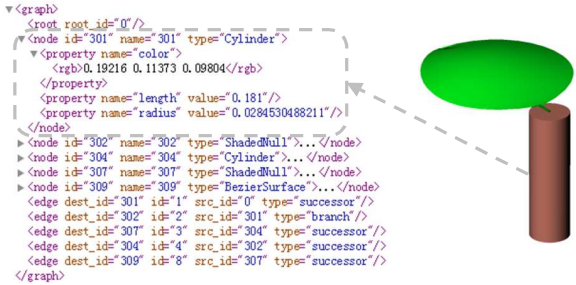

**Figure 5. An example of XEG**

**(left: XEG code, right: corresponding virtual plant).**

We developed an infrastructure based on the specific platforms and models following the proposed architecture. Specific techniques for FSPMs integration have been investigated and applied apart from the techniques commonly used in data integration.

In detail, specific techniques have been used to implement ETL processes for plant information, particularly for geometric data and multiscale structure transformation. For the former, we applied graph traversal in the extracting and loading processes since FSP data models are graphs. More specifically, to extract geometric data, i.e., instances of graphic types, organized by the syntactic structure of FSP data models, we implemented an extracting process that extracts type signatures with argument values from the instances of the source FSPM. We also implemented a transformation process to transform geometric type signatures with given argument values. To enable the transformation, we designed a "dictionary" with entries of signatures for all geometric types defined on the source FSPM platform. In the dictionary, each entry returns signatures of geometric types defined on the target FSPM platform. According to the signature transformations, we designed and implemented a specific algorithm to compute the new argument values for each entry of the dictionary. Besides, we implemented a loading process that creates geometric instances from transformed type signatures with corresponding new arguments. In the dictionary, on one hand, one type can have more than one type signature; one the other hand, one geometric object can be translated into different forms with different argument values, e.g., a parallelogram can be "translated" into two triangles or four triangles. Due to the lack of a feasible solution for automatically picking type signature and translating form (i.e. method to transform arguments), the translation entries and corresponding argument-computing algorithms are fixed and predefined according to the involved graphics libraries. For the latter, we designed specific structural decomposition schemes for each kind of composition to allow the nodes at the finest scale, i.e., metamers, received from MAppleT to be decomposed into nodes of defined geometric types in GroIMP. Figure 6 illustrates the decomposition scheme of a composed metamer defined in MAppleT. Based on this scheme, we implemented a sub transformation process for syntactic and semantic transformation of different systems of scales. Accordingly, a sub transformation process removes the added scale structure when the plant information with added scales is sent back to MAppleT. In this sub process, a property upscaling algorithm that aggregates the same data fields of elementary nodes form the same metamer is essential.

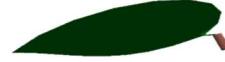

**(a) A plant metamer composed of three transformed shapes: two instances of Cylinder type successively transformed by a translation and an orientation as an internode and a petiole, one instance of BezierPatch type successively transformed by two scaling, an orientation and a translation. Both shape and transformation types are based on the PlantGL library of OpenAlea.**

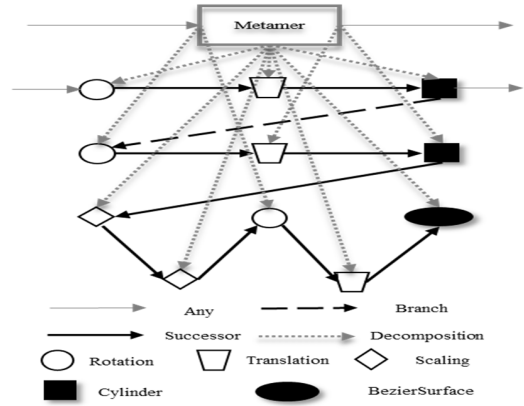

**(b) Decomposition scheme of the metamer (a): elementary geometric instances of types defined on GroIMP are individual nodes in the additional fine scale, and are connected according to the targeting GroIMP model through edges of branch, successor, and decomposition types available on GroIMP.**

**Figure 6. Decompostion scheme (b) for a kind of metamer (a) defined in MAppleT.**

### 3.2 Integrative middleware for GroIMP and OpenAlea based models

Combining all implemented processes, we got the integrative middleware (or an interface) for the integration of specific models based GroIMP and OpenAlea that was required in our project. Its specific parts, which mainly refer to the processes performing transformations of different systems of scales at model level using the decomposition scheme or the property-upscaling algorithm, are only applicable for the integration of the two specific FSPMs. In contrast, the processes at platform level are general and applicable for all FSPMs based on GroIMP and OpenAlea.

We tested our implementation by observing the information coherence. Figure 7 illustrates a non-retroactive test, in which information of an apple tree is generated from MAppleT, and after being processed by the developed middleware, the GroIMP model gets identical information. Figure 8 illustrates a retroactive test, in which a small part of the previous apple tree is first sent from a source model and received by a target model with consistent information, and the source model then modifies the color of the internode to green, and sends the resulting tree information back to the source model. The tree information with the new color has been processed through processes of the implemented middleware in the reverse direction. Therefore, it is possible for the source model to continue with its simulation taking the new color into account.

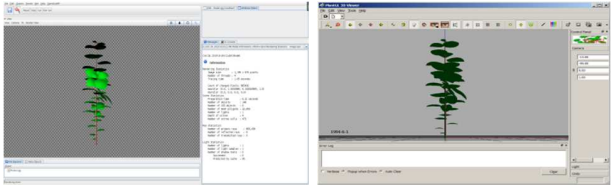

**Figure 7. Plant information is exchanged through the implemented middleware, and the images of the same information are identically visualized on different modeling platforms (left side: GroIMP, right side: OpenAlea).**

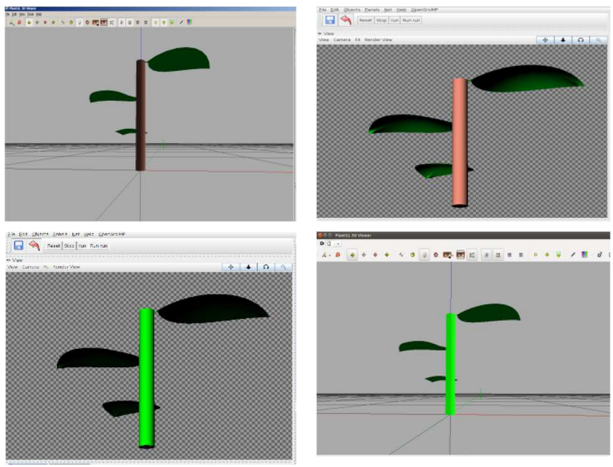

**Figure 8. Plant information is exchanged through the implemented middleware, and the target FSPM modified the color of the internodes to green and sent it back, the modified plant information is identically visualized on different modeling platforms (from upper leaft to lower right: OpenAlea-GroIMP-GroIMP-OpenAlea).**

## 4. CONCLUSION
The designed architecture provides a flexible scheme for the integration of different FSPMs. Its C/S based architecture makes the model integration scalable and allows large complex plant systems modeling. The distinction between processes at platform and model level allows processes at different levels to be developed by specialists in their respective fields (i.e. computer engineers and plant scientists) and the use of developed processes to be optimized and maximized according to their levels.

## 5. ACKNOWLEDGMENT
This work was funded by the German Research Association (DFG) under grant number KU 847/11-1. Gerhard Buck-Sorlin and Faustino Hilario Chi commented on a draft of the manuscript. All support is gratefully acknowledged.